\documentclass[12pt]{elsarticle}

\usepackage{braket}
\usepackage{amsmath}
\usepackage{amsfonts}
\usepackage{color}
\usepackage[dvipsnames]{xcolor}
\usepackage{amssymb}
\usepackage{mathrsfs}
\usepackage{graphicx}
\usepackage{lmodern}
\usepackage{lineno}
\usepackage{hyperref}
\usepackage[normalem]{ulem}
\newcommand{\beq}{\begin{eqnarray}}
\newcommand{\eeq}{\end{eqnarray}}

\newcommand{\nn}{\nonumber}
\newcommand{\tr}{\mathrm{tr}}

\begin{document}

%\linenumbers

%\draftstring{DRAFTv4, 15/12/16}
%\draftfontsize{60}
%%\draftfontfamily{hlh}
%\draftfontfamily{ptm}
%\draftangle{45}
%\definecolor{mycolor}{rgb}{.855,.855,1}
%\draftcolor{mycolor}
%\draftfontattrib{\upshape}

\title{Coherent representation of fields and deformation quantization}

\author[uaslp,dual]{Jasel Berra-Montiel\corref{cor1}}
\ead{jasel.berra@uaslp.mx}
\author[uaslp,dual]{Alberto Molgado}
\ead{alberto.molgado@uaslp.mx}  

\address[uaslp]{Facultad de Ciencias, Universidad Aut\'onoma de San Luis 
Potos\'{\i} \\
Av.~Salvador Nava S/N Zona Universitaria, San 
Luis Potos\'{\i}, SLP, 78290 Mexico}
\address[dual]{Dual CP Institute of High Energy Physics, Mexico}

\cortext[cor1]{Corresponding author}

% \author{Jasel Berra--Montiel$^{1,2}$ and Alberto Molgado$^{1,2}$}
% 
% \address{$^{1}$ Facultad de Ciencias, Universidad Aut\'onoma de San Luis 
% Potos\'{\i} \\
% Campus Pedregal, Av. Parque Chapultepec 1610, Col. Privadas del Pedregal, San
% Luis Potos\'{\i}, SLP, 78217, Mexico}
% \address{$^2$ Dual CP Institute of High Energy Physics, Mexico}

% \eads{\mailto{\textcolor{blue}{jasel.berra@uaslp.mx}},\ 
% \mailto{\textcolor{blue}{alberto.molgado@uaslp.mx}}\ 
% }

% \received{20 October 2016}
% \vspace{-2ex}
% \accepted{}
% \vspace{-2ex}
% \published{}

\begin{abstract}
Motivated by some well known results in the phase space description of quantum optics 
and quantum information theory, we aim to describe the formalism of 
quantum field theory by explicitly considering the holomorphic representation for a scalar field within the deformation quantization 
program.  Notably, the symbol of a symmetric ordered operator in terms of holomorphic variables may be straightforwardly obtained by the quantum field analogue of the Husimi distribution
associated with a normal ordered operator.  This relation also allows to establish a $c$-equivalence between the Moyal and the normal star-products. In addition, by 
writing the density operator in terms of coherent states we are able to directly introduce 
a series representation of the Wigner functional distribution which may be convenient in order to 
calculate  probability distributions of quantum field observables without performing 
formal phase space  integrals at all.

\end{abstract}


% \keywords{Deformation quantization, star-product}
% \ams{81S30, 53D55, 70H45}
%\pacs{}
%\ams{81S30, 53D55, 70H45}

\maketitle

\section{Introduction}

Deformation quantization, also referred to as phase space quantum mechanics by many authors, consists in a general quantization procedure based on the idea that a quantum system is obtained by deforming both the geometrical and the algebraic structures of  classical phase space \cite{Bayen}, \cite{Flato}. One of the main features of this approach lies in the primordial role acquired by the algebra of quantum observables which, from this perspective, is not given by a family of self-adjoint operators acting on a Hilbert space as in ordinary quantum mechanics, but instead these observables correspond to real or complex valued functions defined on the classical phase space, where the usual commutative point-wise product is replaced by an associative, non-commutative product, denoted as the star-product. Consequently, this star-product induces a deformation  of the Poisson bracket in such a way that all the information included on the commutators between self-adjoint operators is contained in the deformed algebraic classical structure. Considering that one may circumvent the use of operators as quantum observables, within the deformation quantization formalism an essential ingredient  resides on the definition of the Wigner distribution \cite{Wigner}. This distribution corresponds to a phase space representation of the density matrix, and thus it contains all the information of the auto-correlation properties and the transition amplitudes associated to a given quantum mechanical system \cite{Zachos}. A prominent property of the Wigner distribution for a quantum system lies on the possibility to acquire negative values on particular regions of phase space and, therefore, it can not be merely interpreted as a probability density in the standard sense and, in consequence, it is usually referred to as a quasi-probability distribution in the literature. However, this seemingly odd feature impress the Wigner distribution with a relevant property since it allows to visualize the fact that quantum trajectories in phase space that correspond to regions with negative probability values may be useful to determine joint-correlation functions and entanglement properties within the quantum system \cite{Weinbub}.

The formalism of deformation quantization not only has provided important contributions in pure mathematics \cite{Kontsevich}, \cite{Reichert}, \cite{Bahns}, \cite{Waldmann}, but it also has proved to be a reliable tool in the understanding of many physical quantum systems \cite{Zachos}, \cite{Fredenhagen}, \cite{Compean}, \cite{Compean2}, recently including the treatment of constrained systems \cite{DQconstraints}, and certain aspects on loop quantum cosmology and quantum gravity \cite{DQpoly}, \cite{PolyWigner}.

In this paper we propose to analyze the formalism of quantum field theory by means of coherent states as seen from the deformation quantization point of view. Motivated by the representation of quantum mechanics in phase space in terms of quasiprobability distributions, widely used in quantum optics and quantum information theory \cite{Cahill}, \cite{Manko}, \cite{Scully}, we address the  construction of the quantum field analogues of the Weyl-Stratonovich quantizer, the Wigner functional, the star-product and the $s$-ordered symbols associated to quantum field operators by adopting the holomorphic representation, as first introduced by Cahill and Glauber in the context of standard quantum mechanics \cite{Cahill}. In particular, via the quantum field analogue of the Husimi distribution we obtain the Moyal and the normal star-products for a scalar field in terms of holomorphic variables \cite{Dito}. Then, following some ideas formulated in \cite{Wunsche}, \cite{Moya}, we provide a series representation of the Wigner functional with respect to the Fock basis and coherent states.

This paper is organized as follows, in section 2, we briefly introduce the basic ideas of the deformation quantization scheme applied to fields and we obtain its holomorphic representation. In section 3, the coherent representation, the star products and the quantum field counterparts of the Wigner and Husimi distributions are obtained. Then, by employing the coherent states, a series representation of the Wigner functional is reviewed. Finally, we introduce some concluding remarks in section 4.         

\section{Deformation quantization of fields }
\label{sec:DQF}

In this section we briefly review the Wigner-Weyl quantization scheme for fields. We will closely follow the description of the formalism as described in \cite{Compean} and \cite{PolyWigner}. For simplicity, we will focus on the real scalar field, nevertheless a generalization to other fields follows straightforwardly. In order to get a full perspective on the deformation quantization program we refer the reader to the more detailed reviews \cite{Bordemann}, \cite{Blaszak}, \cite{Gutt}.  

\subsection{The Wigner-Weyl-Stratonovich quantization}
Consider a real scalar field $\varphi$ defined on a four dimensional background Minkowski spacetime $\mathcal{M}$. Let us perform a $3+1$ decomposition of the spacetime $\mathcal{M}$ in the form $\mathcal{M}=\Sigma\times\mathbb{R}$, for any Cauchy surface $\Sigma$, which at the present case of study may be thought of as topologically equivalent to $\mathbb{R}^{3}$. The spacetime manifold $\mathcal{M}$ is endowed with local coordinates $(x,t)\in\mathbb{R}^{3}\times\mathbb{R}$ and a metric $\eta=diag(+1,+1,+1,-1)$. In what follows, we deal with fields evaluated at the instant $t=0$, and we write $\varphi(x,0):=\varphi(x)$, and $\varpi(x,0):=\varpi(x)$, where $\varpi(x)$ stands for the canonical conjugate momentum associated to the field $\varphi(x)$. Consequently, the phase space of the theory is locally written through coordinates  $\Gamma=(\varphi,\varpi)$, which in turn can be related to initial data on a Cauchy surface $\Sigma$.
    
By analogy to quantum mechanics, in order to define the Weyl quantization rule \cite{Weyl}, i.e., a one to one mapping from the set of classical observables to the set of quantum observables, usually given by self-adjoint operators defined on a Hilbert space, we need to provide a quantization map such that it takes the Poisson brackets
\begin{eqnarray}\label{Poisson}
\left\lbrace \varphi(x),\varpi(y)\right\rbrace 
& = & \delta(x-y) \,,\nn\\
 \left\lbrace \varphi(x),\varphi(y)\right\rbrace 
 & = & 
 \left\lbrace \varpi(x),\varpi(y)\right\rbrace = 0 \,,
\end{eqnarray}  
to the commutator of operators (in natural units where $\hbar=1$)
\begin{eqnarray}
\left[\hat{\varphi}(x),\hat{\varpi}(y) \right]\Psi
& = & 
i\delta(x-y)\Psi \,, \nonumber\\
\left[\hat{\varphi}(x),\hat{\varphi}(y) \right]\Psi
& = & \left[\hat{\varpi}(x),\hat{\varpi}(y) \right]\Psi=0 \,, 
\end{eqnarray}
where the vector state $\Psi[\varphi]$ is given by a functional of the field $\varphi\in\mathcal{S}'(\mathbb{R}^{3})$, where $\mathcal{S}'(\mathbb{R}^{3})$ stands for the Schwartz space of tempered distributions, that is, the space of continuous linear functionals on the space of rapidly decreasing smooth test functions $\mathcal{S}(\mathbb{R}^{3})$. To be more precise, the state $\Psi$ belongs to the Hilbert space $\mathcal{H}=L^{2}\left( \mathcal{S}'(\mathbb{R}^{3}),d\mu\right) $.  Here d$\mu$ is given by the formal measure $\mathcal{D}\varphi:=\prod_{x\in\mathbb{R}^{3}}d\varphi(x)$, which denotes a uniform Lebesgue measure on the configuration space  \cite{Glimm}. Although, it is known that in an infinite dimensional vector space a translational invariant measure cannot be properly defined, other measures such as Gaussian measures of mean zero, can be used in order to construct different representations of the Hilbert space $\mathcal{H}$ by employing proper probability measures, as shown in \cite{PolyWigner}, \cite{Corichi}, \cite{Velinho},.       

The Weyl-Stratonovich quantization, $W:L^{2}(\Gamma)\to \mathcal{L}(\mathcal{H})$, is defined as the linear map from the space of functionals on the phase space $\Gamma$, to the linear operators acting on the Hilbert space $\mathcal{H}$ \cite{Compean}, \cite{Stratonovich}, by 
\begin{equation}\label{Weylmap}
\hat{F}\Psi:=W(F[\varphi,\varpi])\Psi=\int\mathcal{D}\varphi\mathcal{D}\left(\frac{\varpi}{2\pi}\right)F[\varphi,\varpi]\hat{\Omega}[\varphi,\varpi]\Psi \,, 
\end{equation}      
where the operator $\hat{\Omega}$ is given by
\begin{equation}\label{Stratonovich}
\hat{\Omega}[\varphi,\varpi]=\int\mathcal{D}\left(\frac{\lambda}{2\pi} \right)\mathcal{D}\mu\;e^{-i\varphi(\lambda)+i\hat{\varphi}(\lambda)-i\varpi(\mu)+i\hat{\varpi}(\mu)} \,, 
\end{equation}
and we have introduced the notation 
\begin{equation}
\varphi(\lambda):=\int_{\mathbb{R}^{3}}dx\;\varphi(x)\lambda(x)\,,\;\;\;\mathrm{and}\;\;\;\;\hat{\varphi}(\lambda):=\int_{\mathbb{R}^{3}}dx\;\hat{\varphi}(x)\lambda(x) \,,
\end{equation}
for all $\lambda, \mu\in \mathcal{S}(\mathbb{R}^{3})$, and we have used similar expressions to denote the other terms appearing in the definition of $\hat{\Omega}[\varphi,\varpi]$. The operator $\hat{\Omega}$ represented in (\ref{Stratonovich})  corresponds to the quantum field analogue of the Weyl-Stratonovich quantizer, and as we shall see, its inverse allows us to define an associative and noncommutative product, the so called star-product, which proves to be one of the fundamental algebraic structures within the deformation quantization scheme. Employing the relations $\int\mathcal{D}\varphi\ket{\varphi}\bra{\varphi}=\hat{1}$ and $\int\mathcal{D}\left( \frac{\varpi}{2\pi}\right)\ket{\varpi}\bra{\varpi}=\hat{1}$, where $\ket{\varphi}$ and $\ket{\varpi}$ stand for vector eigenstates of the operators $\hat{\varphi}$ and $\hat{\varpi}$, respectively, the Weyl-Stratonovich operator satisfy the following properties
\begin{eqnarray}
\hat{\Omega}^{\dagger}[\varphi,\varpi]
& = & 
\hat{\Omega}[\varphi,\varpi] \,,\\
\tr{\left\lbrace\hat{\Omega}^{\dagger}[\varphi,\varpi]\right\rbrace} 
& = &  1 \,, \\
\tr{\left\lbrace \hat{\Omega}[\varphi,\varpi]\hat{\Omega}[\varphi',\varpi']\right\rbrace}
& = & 
 \delta(\varphi-\varphi')\delta\left( \frac{\varpi-\varpi'}{2\pi}\right) \label{trace} \,.   
\end{eqnarray}
By multiplying the expression (\ref{Weylmap}) by the operator $\hat{\Omega}$ and taking the trace on both sides, using property (\ref{trace}), one obtains that the phase space function associated to the operator $\hat{F}$ reads
\begin{equation}
F[\varphi,\varpi]:=W^{-1}(\hat{F})=\tr\left\lbrace \hat{\Omega}[\varphi,\varpi]\hat{F}\right\rbrace.  
\end{equation}  
This map, also known as the Wigner map, corresponds to the inverse relation of the Weyl-Stratonovich quantizer and it associates quantum operators to real functions or symbols, following the standard terminology in harmonic analysis \cite{Reed}. The symbols determine an associative algebra endowed with a noncommutative product denominated as the Moyal star-product in field theory, given by
\begin{equation}\label{star}
(F_{1}\star F_{2})[\varphi,\varpi]:=W^{-1}(\hat{F}_{1}\hat{F}_{2})=\tr\left\lbrace \hat{\Omega}[\varphi,\varpi]\hat{F}_{1}\hat{F}_{2}\right\rbrace \,.
\end{equation}
By substituting the expression (\ref{Weylmap}) into (\ref{star}) and using the expansion of the functionals $F_{1}$ and $F_{2}$ in Taylor series we obtain (we encourage the reader to see \cite{Compean}, \cite{Hirshfeld} for detailed calculations)
\beq
\label{Moyal}
\hspace{10ex}\mkern-95mu(F_{1}\star F_{2})[\varphi,\varpi]
=  & & 
\nn\\  
F_{1}[\varphi,\varpi]\exp\left\lbrace\frac{i}{2}\int_{\mathbb{R}^{3}} dx \left(\frac{\overleftarrow{\delta}}{\delta\varphi(x)}\frac{\overrightarrow{\delta}}{\delta\varpi(x)}-\frac{\overleftarrow{\delta}}{\delta\varpi(x)}\frac{\overrightarrow{\delta}}{\delta\varphi(x)} \right) \right\rbrace F_{2}[\varphi,\varpi] & . &
\eeq
Finally, let $\hat{\rho}=\ket{\Psi}\bra{\Psi}$ be the density operator of a quantum state $\Psi\in\mathcal{H}$, the symbol $\rho[\varphi,\varpi]$ corresponding to the operator $\hat{\rho}$ reads
\begin{equation}\label{Wignerfunctional}
\rho[\varphi,\varpi]=\tr\left\lbrace \hat{\Omega}[\varphi,\varpi]\hat{\rho}\right\rbrace=\int\mathcal{D}\left(\frac{\mu}{2\pi}\right)e^{-i\varpi(\mu)}\Psi^{*}\left[\varphi-\frac{\mu}{2}\right]\Psi\left[\varphi+\frac{\mu}{2} \right] \,. 
\end{equation} 
This functional is the quantum field analogue of the Wigner distribution associated to the Hilbert space $\mathcal{H}$ in quantum mechanics. 
As mentioned before, a distinguished property for the Wigner distribution lies on the possibility to acquire negative values on some regions of phase space, therefore, it can not be interpreted merely as a probability distribution, and thus it is usually referred to as a quasi-probability distribution. This particularity endows the Wigner distribution with the quality of being an excellent candidate to 
characterize the quantum properties of a system, not only as it contains the information of the density matrix, but also as negative contributions to the Wigner function may be interpreted as non classical interferences, despite the fact that not all quantum states introduce negative contributions to the Wigner distribution \cite{Zachos}.   

\subsection{The Bargamnn-Fock representation}

In order to obtain the Bargmann-Fock representation (also known as the holomorphic representation), let us consider the expansion of the fields $\varphi$ and $\varpi$ in terms of an infinite set of harmonic oscillators
\begin{eqnarray}
\varphi(x,t)=\frac{1}{(2\pi)^{3/2}}\int_{\mathbb{R}^{3}} dk\left(\frac{1}{2\omega(k)}\right)^{1/2}\left(a(k,t)e^{ikx}+a^{*}(k,t)e^{-ikx} \right) \,,\label{varphi} \\
\varpi(x,t)=\frac{1}{(2\pi)^{3/2}}\int_{\mathbb{R}^{3}} dk\left(\frac{\omega(k)}{2}\right)^{1/2}i\left(a^{*}(k,t)e^{-ikx}-a(k,t)e^{ikx} \right)\label{varpi} \,, 
\end{eqnarray}
where $\omega(k)=\sqrt{k^{2}+m^{2}}$, $a(k,t)=a(k)e^{-i\omega(k)t}$ and 
$kx:=\sum_{i=1}^3 k_{i}x_{i}$, with $i=1,2,3$. From expressions (\ref{varphi}) and (\ref{varpi}) we may obtain
\begin{equation}\label{a}
a(k,t)=\frac{1}{(2\pi)^{3/2}(2\omega(k))^{1/2}}\int dx\, e^{-ikx}\left(\omega(k)\varphi(x,t)+i\varpi(x,t) \right) \,,  
\end{equation}
and its complex conjugate, $a^{*}(k,t)$. By using the Poisson bracket relations (\ref{Poisson}), we can observe that $a(k,t)$ and $a^{*}(k,t)$ satisfy the Poisson algebra
\begin{eqnarray}\label{Poissona}
\left\lbrace a(k,t),a^{*}(k',t)\right\rbrace
& = & 
-i\delta(k-k') \,,\\
\left\lbrace a(k,t),a(k',t)\right\rbrace
& = & 
\left\lbrace a^{*}(k,t),a^{*}(k',t)\right\rbrace=0\label{Poissonas} \,.
\end{eqnarray}
In order to compute the Weyl-Stratonovich quantizer in terms of the holomorphic variables $a(k,t)$ and $a^{*}(k,t)$, let us introduce the following canonical variables \cite{Lifshitz}
\begin{eqnarray}
Q(k,t)=\left( \frac{1}{2\omega(k)}\right)^{1/2}\left(a(k,t)+a^{*}(k,t) \right)\,,\label{Q}\\
P(k,t)=i\left(\frac{\omega(k)}{2} \right)^{1/2}\left(a^{*}(k,t)-a(k,t) \right)\,. \label{P}   
\end{eqnarray}
From the Poisson algebra given in (\ref{Poissona}) and (\ref{Poissonas}), and the definition of the variables (\ref{Q}) and (\ref{P}), one can derive the Poisson relations
\begin{eqnarray}
\left\lbrace Q(k,t),P(k',t)\right\rbrace
& = & \delta(k-k') \,,\label{PoissonQP}\\
\left\lbrace Q(k,t),Q(k',t)\right\rbrace
& = & 
\left\lbrace P(k,t),P(k',t)\right\rbrace=0\label{PoissonQQ} \,.
\end{eqnarray}
Further, whenever we insert the holomorphic variables (\ref{a}) into equations (\ref{Q}) and (\ref{P}) we get
\begin{eqnarray}
\nn
Q(k,t)
& = & 
\frac{1}{(2\pi)^{3/2}\omega(k)}\int_{\mathbb{R}^{3}} dx\left(\varpi(x,t)\sin(kx)+\omega(k)\varphi(x,t)\cos(kx) \right) \,,\\
P(k,t)
& = & \frac{1}{(2\pi)^{3/2}}\int_{\mathbb{R}^{3}} dx\left( \varpi(x,t)\cos(kx)-\omega(k)\varphi(x,t)\sin(kx)\right)\,.  
\label{Q&P}
\end{eqnarray} 
Then, it is evident from the algebraic relations (\ref{PoissonQP}), (\ref{PoissonQQ}) and (\ref{Q&P}) that $Q(k,t)$ and $P(k,t)$ defines a linear canonical transformation. This means, that instead of using the field variables $\varphi(x)$ and $\varpi(x)$, it should be possible to write the Weyl-Stratonovich quantizer in terms of $Q(k,t)$ and $P(k,t)$. 
With the computation of the canonical variables $Q(k,t)$ and $P(k,t)$ in hand, we can rewrite the Weyl-Stratonovich map in the following form (see \cite{Compean} for further details)
\begin{equation}\label{WSQP}
\hat{\Omega}[Q,P]=\hat{\Omega}[\varphi[Q,P],\varpi[Q,P]]=\int\mathcal{D}\left(\frac{\lambda}{2\pi}\right)\mathcal{D}\mu\, e^{-iQ(\lambda)-iP(\mu)+i\hat{Q}(\lambda)+\hat{P}(\mu)} \,,  
\end{equation}
where we have used the notation 
\begin{equation}
Q(\lambda)=\int_{\mathbb{R}^{3}}dk\;Q(k)\lambda(k)\,, \;\;\; \mathrm{and} \;\;\;\; \hat{Q}(\lambda)=\int_{\mathbb{R}^{3}}dk\;\hat{Q}(k)\lambda(k)\,,
\end{equation}
for all $\lambda, \mu\in \mathcal{S}(\mathbb{R}^{3})$, and similar expressions to denote the other terms appearing in the definition of $\hat{\Omega}[Q,P]$. The field operators $\hat{Q}$ and $\hat{P}$ satisfy 
the relations $\hat{Q}(k)\ket{Q}=Q(k)\ket{Q}$, $\hat{P}(k)\ket{P}=P(k)\ket{P}$, where $\ket{Q}$ and $\ket{P}$ stand for their corresponding eigenvectors fulfilling the properties $\int DQ \ket{Q}\bra{Q}=\hat{1}$ and 
$\int \mathcal{D}\left( \frac{P}{2\pi}\right)\ket{P}\bra{P}=\hat{1}$, respectively.
Consequently, it can be shown that the Moyal star-product can be written as (see \cite{Compean} for more details)
\begin{eqnarray}
\mkern-110mu(F_{1}\star F_{2})[Q,P]=\tr\left\lbrace \hat{\Omega}[Q,P]\hat{F}_{1}\hat{F}_{2} \right\rbrace = \nonumber\\
F_{1}[Q,P]\exp\left\lbrace\frac{i}{2}\int_{\mathbb{R}^{3}} dk \left(\frac{\overleftarrow{\delta}}{\delta Q}\frac{\overrightarrow{\delta}}{\delta P(k)}-\frac{\overleftarrow{\delta}}{\delta P(k)}\frac{\overrightarrow{\delta}}{\delta Q(k)} \right) \right\rbrace F_{2}[Q,P] \,. 
\end{eqnarray}
Finally, it is possible, using the formalism in terms of the $Q(k)$ and $P(k)$ variables to express the symbol of the density operator $\hat{\rho}=\ket{\Psi}\bra{\Psi}$ as
\begin{equation}
\rho[Q,P]=\tr\left\lbrace \hat{\Omega}[Q,P]\hat{\rho}\right\rbrace=\int\mathcal{D}\left(\frac{\mu}{2\pi}\right)e^{-i P(\mu)}\Psi^{*}\left[Q-\frac{\mu}{2}\right]\Psi\left[Q+\frac{\mu}{2} \right] \,. 
\end{equation} 
In the next section we will find that, by making use of coherent states, the Weyl-Stratonovich quantizer and the Wigner functional written in terms of holomorphic variables will allow us to compute the quantum field analogue of the $s$-ordered symbols introduced within the context of quantum mechanics by Cahill and Glauber in~\cite{Cahill} with the aim to study different orderings in comparison to the Weyl-Wigner symmetric ordering of operators and their corresponding star-products.

\section{Coherent representation and deformation quantization}
\subsection{Coherent representation and star products}

In order to construct coherent states in quantum field theory, i.e., the analogues of well-localized states of the harmonic oscillator with fluctuations similar as those for the vacuum state \cite{Itzykson}, let us assume a normalizable function $\alpha(k)\in\mathcal{S}(\mathbb{R}^{3})$, and consider the state given by
\begin{equation}\label{displacement}
\ket{\alpha}=e^{\hat{a}^{\dagger}(\alpha)-\hat{a}(\alpha^{*})}\ket{0}=:\hat{D}(\alpha)\ket{0} \,, 
\end{equation} 
where, as before, we have used the notation 
\begin{equation}\label{coherent}
\hat{a}^{\dagger}(\alpha)=\int_{\mathbb{R}^{3}}dk\,\alpha(k)\hat{a}^{\dagger}(k)
\,, \;\;\; \mathrm{and}\;\;\;\; \hat{a}(\alpha^{*})=\int_{\mathbb{R}^{3}}dk\;\alpha^{*}(k)\hat{a}(k) \,,
\end{equation}
and $\ket{0}$ stands for the vacuum ground state defined through
\beq\label{vacuum}
\hat{a}(k)\ket{0} = 0 \,, \;\; \mathrm{and} \;\;\;\; \braket{0|0}=1 \,,
\eeq
for all $k\in\mathbb{R}^{3}$, and $\hat{a}^{\dagger}(k)$, $\hat{a}(k)$ denote the creation and annihilation operators for a given mode $k$, respectively. In the definition (\ref{displacement}), the operator $\hat{D}(\alpha)$ corresponds to the field counterpart of the Glauber displacement operator \cite{Cahill}, and it satisfies
the relation\footnote{Throughout the paper, given any complex number $a$, we denote by $\Re (a)$ and $\Im (a)$ its real and imaginary parts, respectively.}
\begin{equation}
\label{DisplProp}
\hat{D}(\alpha)\hat{D}(\beta)\ket{0}=\hat{D}(\alpha)\ket{\beta}=e^{i\Im(\braket{\beta,\alpha})}\ket{\alpha+\beta} \,,
\end{equation} 
where $\braket{\beta,\alpha}$ represents the inner product in $L^{2}(\mathbb{R}^{3})$
\begin{equation}
\braket{\beta,\alpha}=\int_{\mathbb{R}^{3}}dk\,\beta^{*}(k)\alpha(k) \,.
\end{equation}
Moreover, the coherent states satisfy the non-orthogonality relations
\begin{equation}
\braket{\beta|\alpha}=e^{\frac{1}{2}(\braket{\beta,\alpha}-\braket{\alpha,\beta})}e^{-\frac{1}{2}||\beta-\alpha||^{2}} \,,
\end{equation}
where $||\beta-\alpha||^{2}$ denotes the norm induced by the inner product $\braket{\beta-\alpha,\beta-\alpha}$. Even though the coherent states are 
non-orthogonal, one may show that they fulfill the completeness relation \cite{Glimm}, \cite{Itzykson}
\begin{equation}\label{completeness}
\int\mathcal{D}^{2}\left(\frac{\alpha}{\pi}\right)\ket{\alpha}\bra{\alpha}=\hat{1} \,, 
\end{equation}
where the measure is explicitly written as $\mathcal{D}^{2}\left(\frac{\alpha}{\pi}\right)=\mathcal{D}\left(\frac{\Re({\alpha})}{\pi}\right)\mathcal{D}\left(\Im({\alpha})\right)$, that is, it is a formal integral over the complex $\alpha(k)$-plane, for all $k\in\mathbb{R}^{3}$. 

In order to write the Weyl-Stratonovich quantizer in terms of the coherent state formulation of fields, let us substitute equations (\ref{Q}) and (\ref{P}) into the expression (\ref{WSQP}), and by making a change of variables we find
\beq
\mkern-50mu \hat{\Omega}[a,a^{*}]
& =& \int \mathcal{D}^{2}\left(\frac{\xi}{\pi^{2}}\right)e^{a(\xi^{*})-a^{*}(\xi)+\hat{a}^{\dagger}(\xi)-\hat{a}(\xi^{*})}  \nn\\
& = & 
\int \mathcal{D}^{2}\left(\frac{\xi}{\pi^{2}}\right)e^{a(\xi^{*})-a^{*}(\xi)}\hat{D}(\xi) \,,
\label{WSa}
\eeq 
where the complex function $\xi(k)$ is such that $\Re(\xi),\Im(\xi)\in\mathcal{S}(\mathbb{R}^{3})$, and the formal measure is explicitly given by $\mathcal{D}^{2}\left(\frac{\xi}{\pi^{2}}\right)=\mathcal{D}\left(\frac{\Re(\xi)}{\pi}\right)\mathcal{D}\left(\frac{\Im(\xi)}{\pi}\right)$. It is evident now, that the operator $\hat{\Omega}[a,a^{*}]$ defined in (\ref{WSa}) corresponds to the normalized quantum field analogue of the symmetric ordered Weyl-Stratonovich operator employed in quantum optics and quantum information  \cite{Manko}, \cite{Scully}, 
\cite{Marmo}. By using some standard techniques on Gaussian functional integration \cite{Glimm} and the completeness relation (\ref{completeness}), the operator $\hat{\Omega}[a,a^{*}]$ satisfies $\tr\left\lbrace\hat{\Omega}[a,a^{*}]\right\rbrace=1$. 

Now, with the Weyl-Stratonovich operator at hand, the symbol associated to an operator $\hat{F}$ in terms of holomorphic variables is given by
\begin{equation}
F[a,a^{*}]:=W^{-1}[\hat{F}]=\tr\left\lbrace \hat{\Omega}[a,a^{*}]\hat{F} \right\rbrace \,. 
\end{equation}
By using the Baker-Campbell-Hausdorff formula \cite{Hall} and the explicit expression for $\hat{\Omega}[a,a^*]$ in (\ref{WSa}), we can write the symbol of an operator as
\begin{equation}\label{FG}
F[a,a^{*}]=\int \mathcal{D}\left(\frac{\xi}{\pi^{2}}\right) e^{\braket{\xi,a}-\braket{a,\xi}+\frac{1}{2}||\xi||^{2}}G(\xi) \,,
\end{equation}
where
\begin{equation}
G(\xi):=\tr\left\lbrace \hat{D}(\xi)\hat{F}\right\rbrace e^{-\frac{1}{2}||\xi||^{2}} \,.
\end{equation}
From equation (\ref{FG}), we note that the symbol associated to the operator $\hat{F}$, can be written as
\begin{equation}
F[a,a^*]=\sum_{n=0}^{\infty}\frac{1}{2^{n}n!}\int \mathcal{D}\left(\frac{\xi}{\pi^{2}}\right)||\xi||^{2n} e^{\braket{\xi,a}-\braket{a,\xi}}G(\xi) \,.
\end{equation}
To simplify this expression, we note that
\begin{equation}
\int_{\mathbb{R}^{3}} dk \frac{\delta}{\delta a(k)}\frac{\delta}{\delta a^{*}(k)}e^{\braket{\xi,a}-\braket{a,\xi}}=-||\xi||^{2}e^{\braket{\xi,a}-\braket{a,\xi}} \,, 
\end{equation}
then, the symbol $F[a,a^{*}]$ reads
\begin{equation}\label{WH}
F[a,a^{*}]=\exp\left\lbrace  {-\frac{1}{2}\int_{\mathbb{R}^{3}}} dk \frac{\delta}{\delta a(k)}\frac{\delta}{\delta a^{*}(k)}\right\rbrace \int \mathcal{D}\left(\frac{\xi}{\pi^{2}}\right)e^{\braket{\xi,a}-\braket{a,\xi}}G(\xi) \,.
\end{equation}
From this last identity we realize, by comparing with the $s$-parametrized quasiprobability distributions in quantum mechanics \cite{Cahill}, \cite{Moya}, that $F[a,a^{*}]$  is related to the symbol associated to the quantum field analogue of the Husimi $Q$-representation of a normal ordered operator given by 
\begin{equation}
F^{(N)}[a,a^{*}]:= Q^{-1}[\hat{F}]=\int \mathcal{D}\left(\frac{\xi}{\pi^{2}}\right)e^{\braket{\xi,a}-\braket{a,\xi}}G(\xi)  \,.
\end{equation}
Further, by means of the completeness relation (\ref{completeness}) and the functional integration of Gaussians \cite{Glimm}, it is possible to express the symbol for $F^{(N)}[a,a^{*}]$ as
\begin{equation}\label{Husimi}
F^{(N)}[a,a^{*}]=\tr\left\lbrace\hat{F}\hat{\rho}\right\rbrace= \braket{a|\hat{F}|a} \,.
\end{equation}
Finally, we can observe that, within the context of quantum field theory, the symbol of an operator in the symmetric ordering stated by the Weyl-Stratonovich operator is related to the normal ordering given by the Husimi $Q$-representation, so that
\begin{equation}
F[a,a^{*}]=W^{-1}[\hat{F}]=\exp\left\lbrace  {-\frac{1}{2}\int_{\mathbb{R}^{3}}} dk \frac{\delta}{\delta a(k)}\frac{\delta}{\delta a^{*}(k)}\right\rbrace F^{(N)}[a,a^{*}] \,.
\end{equation} 

We now look for the composition rule for symbols, which determines the star product. By definition, the Moyal star-product is given as follows
\begin{equation}
(F_{1}\star F_{2})[a,a^{*}]=\tr\left\lbrace\hat{\Omega}[a,a^{*}]\hat{F}_{1}\hat{F}_{2} \right\rbrace \,, 
\end{equation} 
by applying formulas (\ref{WH}) and (\ref{Husimi}) we obtain
\begin{equation}\label{Moyala}
(F_{1}\star F_{2})[a,a^{*}]=\exp\left\lbrace  {-\frac{1}{2}\int_{\mathbb{R}^{3}}} dk \frac{\delta}{\delta a(k)}\frac{\delta}{\delta a^{*}(k)}\right\rbrace \braket{a|\hat{F}_{1}\hat{F}_{2}|a} \,.
\end{equation}
Since the Husimi $Q$-representation deals with normal ordered operators, in such scheme we can write them as the normal ordered expansion 
\begin{equation}
\hat{F}=\sum_{m,n=0}\int_{\mathbb{R}^{3}}dk\;c_{mn}(k)(\hat{a}^{\dagger}(k))^{m}(\hat{a}(k))^{n} \,,
\end{equation} 
so that, in terms of the expectation value over coherent states and employing the completeness relation (\ref{completeness}), we observe that
\begin{eqnarray}
(F_{1}\star_{N}F_{2})[a,a^{*}] 
:=  \braket{a|\hat{F}_{1}\hat{F}_{2}|a} = \nonumber\\
\sum_{m,n,p,q}\int_{\mathbb{R}^3\times\mathbb{R}^{3}}dkdk'c_{mn}(k)b_{pq}(k')\braket{a|(\hat{a}^{\dagger}(k))^{m}(\hat{a}(k))^{n}(\hat{a}^{\dagger}(k'))^{p}(\hat{a}(k'))^{q}|a} =\nonumber\\
\int\mathcal{D}^{2}\left(\frac{\xi}{\pi}\right)F_{1}[\xi,a^{*}]F_{2}[a,\xi^{*}]e^{-||a-\xi||^{2}} \,.
\end{eqnarray}
As in the case of the Moyal star-product (\ref{Moyal}), we can perform an expansion of the functionals $F_{1}$ and $F_{2}$ in Taylor series \cite{Alexanian}, \cite{Lizzi}, and then by performing the functional Gaussian integral, yields
\begin{equation}
(F_{1}\star_{N}F_{2})[a,a^{*}]=F_{1}[a,a^{*}]\exp\left\lbrace\int_{\mathbb{R}^{3}}dk\left(\frac{\overleftarrow{\delta}}{\delta a(k)}\frac{\overrightarrow{\delta}}{\delta a^{*}(k)} \right) \right\rbrace F_{2}[a,a^{*}] \,.
\end{equation}
This $\star_{N}$ product  corresponds to the normal star-product obtained by means of Berezin calculus in \cite{Dito}. Then, it turns out that from the equation (\ref{Moyala}), the Moyal star-product is given by
\begin{equation}\label{cequivalence}
(F_{1}\star F_{2})[a,a^{*}]=\exp\left\lbrace  {-\frac{1}{2}\int_{\mathbb{R}^{3}}} dk \frac{\delta}{\delta a(k)}\frac{\delta}{\delta a^{*}(k)}\right\rbrace (F_{1}\star_{N}F_{2})[a,a^{*}]\,.
\end{equation}
The relation between the two star-products as depticted in (\ref{cequivalence}) is known as a $c$-equivalence within the formalism of deformation quantization \cite{Dito}, \cite{HirshfeldP}, and by expanding the last exponential we obtain
\beq
\mkern-90mu (F_{1}\star F_{2})[a,a^{*}]=
\nn\\
F_{1}[a,a^{*}]\exp\left\lbrace\frac{1}{2}\int_{\mathbb{R}^{3}}dk\left(\frac{\overleftarrow{\delta}}{\delta a(k)}\frac{\overrightarrow{\delta}}{\delta a^{*}(k)}-\frac{\overleftarrow{\delta}}{\delta a^{*}(k)}\frac{\overrightarrow{\delta}}{\delta a(k)} \right) \right\rbrace F_{2}[a,a^{*}] \,,
\eeq
which corresponds to the star-product obtained in \cite{Dito}.

To conclude this section, and within the deformation quantization scheme for quantum field theories developed in this work, we note that it is possible to generalize the Weyl-Stratonovich quantization map for any $s$-ordered symbol by a straightforwardly comparison with the standard quantum mechanics counterpart, as
explored in~\cite{Cahill}.  Indeed, by considering the s-ordered quantization operator
\begin{equation}
\hat{\Omega}_{s}[a,a^{*}]=\int \mathcal{D}^{2}\left(\frac{\xi}{\pi^{2}}\right)e^{a(\xi^{*})-a^{*}(\xi)}e^{\frac{1}{2}s||\xi||^{2}}\hat{D}(\xi) \,,
\end{equation}  
where $s$ is a continuous parameter which defines the ordering of the creation and annihilation operators. The case $s=0$ corresponds to the standard Weyl-Wigner symmetric ordering, and for $s=-1$ we have the Husimi normal ordering, both cases analyzed above. An additional relevant value corresponds to $s=1$, which yields the Glauber-Sudarshan anti-normal ordering in quantization \cite{Cahill}, \cite{Scully}. The importance of the $s$-parametrization as a useful tool for understanding the convergence properties of the expansion of operator functions in terms of integral representations and its corresponding integral kernels has been characterized in \cite{Soloviev}. 

\subsection{Series representation of the Wigner functional}

It is possible to obtain a series representation of the Wigner functional $\rho[\varphi,\varpi]$ in complete analogy to the Wigner function appearing in quantum mechanics \cite{Wunsche}, \cite{Moya}. By making the substitution of $\gamma=-\mu/2$ in the definition (\ref{Wignerfunctional}), we obtain
\begin{equation}
\rho[\varphi,\varpi]=\int\mathcal{D}\left( \frac{\gamma}{\pi}\right)e^{2i\varpi(\gamma)}\braket{\varphi-\gamma|\hat{\rho}|\varphi+\gamma} \,, 
\end{equation}
where $\hat{\rho}=\ket{\Psi}\bra{\Psi}$ is the density operator associated to a quantum state $\Psi\in\mathcal{H}$. Then
\begin{eqnarray}
\rho[\varphi,\varpi]&=&\int\mathcal{D}\left( \frac{\gamma}{\pi}\right)e^{2i\varpi(\gamma)}\braket{-\gamma|e^{-i\hat{\varpi}(\varphi)}\hat{\rho}e^{i\hat{\varpi}(\varphi)}|\gamma}  \nn \\
&=&\int\mathcal{D}\left( \frac{\gamma}{\pi}\right)\braket{-\gamma|e^{-i\hat{\varphi}(\varpi)}e^{-i\hat{\varpi}(\varphi)}\hat{\rho}e^{i\hat{\varpi}(\varphi)}e^{i\hat{\varphi}(\varpi)}|\gamma} \,,
\end{eqnarray} 
by using the Baker-Campbell-Hausdorff formula and introducing the parity operator $\hat\Pi\ket{\gamma}=\ket{-\gamma}$, it follows that
\begin{eqnarray}
\rho[\varphi,\varpi]&=&\int\mathcal{D}\left( \frac{\gamma}{\pi}\right)\braket{\gamma|\hat{\Pi}e^{-i\hat{\varphi}(\varpi)}e^{-i\hat{\varpi}(\varphi)}\hat{\rho}e^{i\hat{\varpi}(\varphi)}e^{i\hat{\varphi}(\varpi)}|\gamma}  \,\nn\\
&=&\tr{\left\lbrace \hat{\Pi}\hat{D}^\dagger(\xi)\hat{\rho}\hat{D}(\xi)\right\rbrace } \,,
\end{eqnarray}
where $\hat{D}(\xi)$ is the displacement operator defined in (\ref{displacement}) and $\xi\in\mathcal{S}(\mathbb{R}^{3})$ depends on the field variables $\varphi$ and $\varpi$. From now on, we will consider the Fock basis \cite{Zeidler} spanned
by vector states with occupation numbers $n_{s}$ for all modes $k$
\begin{equation}\label{Fock}
\ket{n_{1},n_{2},\ldots}=\left[ \prod_{s}\frac{\left( \hat{a}^{\dagger}(k_{s})\right)^{n_{s}}}{\sqrt{n_{s}!}}\right]\ket{0} \,,  
\end{equation}
that is, the vector (\ref{Fock}) corresponds to the quantum state in which $n_{1}$ particles have momentum $k_{1}$, $n_{2}$ particles have momentum $k_{2}$, etc., and $\ket{0}$ denotes the vacuum ground state defined in (\ref{vacuum}) above.
The set of state vectors $\ket{n_{1},n_{2},\ldots}$, with all possible choices of $n_{s}$,  form a complete orthonormal basis in the Hilbert space. Thus, we finally obtain the following representation of the Wigner functional
\begin{equation}\label{Wseries}
\rho[\varphi,\varpi]=\sum_{n_{1},n_{2},\ldots}\braket{n_{1},n_{2},\ldots|\hat{\Pi}\hat{D}^\dagger(\xi)\hat{\rho}\hat{D}(\xi)|n_{1},n_{2},\ldots} \,.
\end{equation}
By writing the density operator in terms of coherent states, $\hat{\rho}=\ket{\alpha}\bra{\alpha}$, and recalling the definition of the displacement operators (\ref{displacement}) and using the property~(\ref{DisplProp}), we can express the formula (\ref{Wseries}) as
\begin{equation}
\rho[\varphi,\varpi]=\sum_{n_{1},n_{2},\ldots}\braket{n_{1},n_{2},\ldots|\hat{\Pi}\hat{D}^\dagger(\xi)\hat{D}(\alpha)\ket{0}\bra{0}\hat{D}^{\dagger}(\alpha)\hat{D}(\xi)|n_{1},n_{2},\ldots} \,.
\end{equation}
Further, using the fact that \cite{Scully},
\begin{equation}
\hat{D}^{\dagger}(\xi)\hat{D}(\alpha)\ket{0}=e^{\frac{1}{2}(\braket{\xi,\alpha}-\braket{\alpha,\xi})}\ket{\alpha-\xi} \,,
\end{equation}
and expressing the parity operator as $\hat{\Pi}=(-1)^{\hat{N}}$, where $\hat{N}$ corresponds to the number operator in field theory \cite{Zeidler}, $\hat{N}=\frac{1}{(2\pi)^{3}}\int_{\mathbb{R}^{3}}dk\,\hat{a}^{\dagger}(k)\hat{a}(k)$,
the Wigner functional reads
\begin{equation}\label{Wseriescoherent}
\rho[\varphi,\varpi]=\sum_{n_{1},n_{2},\ldots}(-1)^N\braket{n_{1},n_{2},\ldots|\alpha-\xi}\braket{\alpha-\xi|n_{1},n_{2},\ldots} \,.
\end{equation}
The series representation of the Wigner functional contained in (\ref{Wseries}) and (\ref{Wseriescoherent}) allow us to calculate the probability distribution of quantum field observables by avoiding formal phase space integrals. This approach could be useful in circumstances where integration methods are very challenging as it may be the case, for instance, while studying either quantum effects in curved backgrounds or in quantum gravity \cite{Birrell}.   

\section{Conclusions}
\label{sec:conclu} 
 
In this paper we have analyzed  the formalism of quantum field theory by means of coherent states appropriately introduced within the context of deformation quantization.
In particular,  by selecting the holomorphic representation for a scalar field, we have explicitly encountered  the quantum field analogues of the Weyl-Stratonovich quantizer, the Wigner distribution and the Moyal star-product by means of the Husimi distribution that corresponds to normal ordered operators.  By construction, we also have set
a $c$-equivalence between the Moyal product and the normal star-product, $\star_{N}$, 
which is commonly obtained by applying Berezin calculus. 
Further, within our context, we have also noted that the correspondence between the symmetric and normal ordered symbols  may be enlarged by considering the extension of the Weyl-Stratonovich quantization map for any $s$-ordered symbol in terms of the Glauber 
displacement operator and a real parameter $s$.   This may be a very relevant issue as the $s$-parametrization 
has been proposed in the literature as a useful tool in order to analyze convergence of the expansion of operator functions in terms of integral representations. 

Moreover, we have discussed a series representation of the Wigner functional which is obtained by considering the action of the Glauber displacement operators on the density operator written in terms of coherent states associated to quantum fields.   This series representation may demonstrate relevant in order to determine the probability distribution of quantum field observables.  We expect this will be the case in quantum field theory in a curved background where cumbersome phase space integrals are involved even in the simpler cases.  
From a different perspective, it will also be relevant to implement the techniques described here within the context of either gauge field theories or the non-regular quantum representations that naturally emerge in Loop Quantum Gravity (LQG).  Indeed, we expect that our results here will be significant, in addition to the developments in~\cite{DQconstraints}, \cite{DQpoly} and \cite{PolyWigner}, in order to explore the LQG coherent state formulation in the cosmological scenario. Work along these lines is in progress, and will be reported elsewhere.

\section*{Acknowledgments}
The authors would like to acknowledge financial support from CONACYT-Mexico
under project CB-2017-283838.

% \section*{References}

\bibliographystyle{unsrt}

\end{document}